\documentstyle[emulateapj,psfig]{article}

\newcommand{\etal}	{et~al.\ }
\newcommand{\ammonia}	{\mbox{NH$_3$}}
\newcommand{\pamm}	{\mbox{NH$_4^+$}}
\newcommand{\damm}	{\mbox{NH$_2$D}}
\newcommand{\pdamm}	{\mbox{NH$_3$D$^+$}}
\newcommand{\ddamm}	{\mbox{NHD$_2$}}
\newcommand{\tdamm}	{\mbox{ND$_3$}}

\newcommand{\nndp}	{\mbox{N$_2$D$^+$}}
\newcommand{\dcop}	{\mbox{DCO$^+$}}
\newcommand{\gamh}	{\mbox{$\Gamma_{\mbox{\tiny H$^+$}}$}}
\newcommand{\gamd}	{\mbox{$\Gamma_{\mbox{\tiny D$^+$}}$}}

\begin{document}

\slugcomment{To appear in ApJ vol 553}

\title{Gas phase production of NHD$_2$ in L134N}

\author{S. D. Rodgers and S. B. Charnley}
\affil{Space Science Division, MS 245-3, NASA Ames Research
Center, Moffett Field, CA 94035}

\begin{abstract}
We show analytically that large abundances of \damm\ and \ddamm\ can
be produced by gas phase chemistry in the interiors of cold dense
clouds.  The calculated fractionation ratios are in good agreement
with the values that have been previously determined in L134N and
suggest that triply-deuterated ammonia could be detectable in dark
clouds. Grain surface reactions may lead to similar \damm\ and \ddamm\
enhancements but, we argue, are unlikely to contribute to the
deuteration observed in L134N\@.
\end{abstract}
\keywords{ISM: individual objects: L134N -- ISM: chemistry -- ISM: deuterium}

\section{Introduction}

Recently \ddamm\ and \damm\ were detected in L134N by Roueff \etal
(2000). This is the first detection of a doubly-deuterated molecule in
a dark interstellar cloud. Roueff \etal derived fractionation ratios
of $R(\damm)=0.1$ and $R(\ddamm)=0.05$, where we define the
fractionation ratio, $R$, of the deuterated species XH$_m$D$_{m'}$ to
be $n$(XH$_m$D$_{m'}$)/$n$(XH$_{m+1}$D$_{m'-1}$) where $n$ is the
number density. \damm\ was previously detected at the same position by
Olberg \etal (1985) and Saito \etal (2000), who determined
$R(\damm)\approx 0.05$. The first interstellar detection of \nndp\ was
also made at approximately the same location by Snyder \etal (1977),
who derived $R(\nndp) = 0.45$. \dcop\ has been detected in L134N by
Wootten, Loren, \& Snell (1982), Gu\'elin, Langer, \& Wilson (1982),
and Butner, Lada, \& Loren (1995) with values for $R(\dcop)$ in the
range 0.03--0.07. More recent observations of \nndp\ and \dcop\ at the
same location as the \ddamm\ peak were carried out by Tin\'e \etal
(2000), who obtained fractionation ratios of 0.35 and 0.18
respectively for these ions. In addition to the deuterium
fractionation, it is known that the absolute abundances relative to
H$_2$ of \ammonia\ and N$_2$H$^+$ peak at the same position
(Ungerechts, Walmsley, \& Winnewisser 1980; Swade 1989; Dickens \etal
2000).

The only other doubly deuterated molecule observed in the interstellar
medium is D$_2$CO, which has been detected in two regions: the
well-known Orion Compact Ridge source (Turner 1990) and the low-mass
protostar IRAS 16293-2422 (Ceccarelli \etal 1998; Loinard \etal
2000). In both cases, the large D$_2$CO fractionation -- $R$(D$_2$CO$)
= 0.02$ and 0.35 respectively -- is thought to result from grain
surface chemistry, since the molecular abundances in these warm
regions reflect the evaporation of ice mantles from interstellar dust
grains (e.g.\ Brown, Charnley, \& Millar 1988), and the
post-evaporation molecular D/H ratios remain equal to the ratios in
the precursor ices for over $10^4$ years (Rodgers \& Millar 1996).
Deuterated formaldehyde forms on grains via addition of H and D atoms
to CO (Tielens 1983; Charnley, Tielens, \& Rodgers 1997), and so large
amounts of HDCO and D$_2$CO can be expected if the gas-phase atomic
D/H ratio is large.

\damm\ and \ddamm\ can also form on grains through D and H atom
additions to atomic N (Brown \& Millar 1989). However, as noted by
Roueff \etal (2000), the lack of ongoing star formation, together with
the low temperature in L134N ($T\approx 9$--12\,K; Swade 1989; Dickens
\etal 2000), suggests that mantle removal has not occurred.  Markwick,
Millar, \& Charnley (2000) have proposed a nonthermal mechanism for
removing grain mantles in dark clouds involving ion-neutral streaming
produced in MHD motions.  However, this process also acts to remove
$\rm H_2D^+$ and $\rm N_2D^+$ from the gas (Charnley 1998) and so this
mechanism is unlikely to have occurred at the deuterium emission peak
in L134N\@. It therefore appears that the \damm\ and \ddamm\ observed
in L134N must be created in the gas phase.

In this paper, we address the issue of gas phase ammonia deuteration.
We first discuss the underlying chemistry which controls the deuterium
fractionation and ammonia abundances in dark clouds. We then show how
successive deuteron transfer reactions can lead to large abundances of
multiply-deuterated ammonia. We derive analytical expressions for the
steady-state D/H ratios in isotopomers of ammonia, and compare these
theoretical ratios with those observed in L134N and with those
expected from grain surface chemistry.


\section{Deuterium Fractionation in Dark Clouds}

A number of singly-deuterated molecules have been detected in dark
interstellar clouds with large fractionations ($R\sim0.001$--0.1), and
the theory behind the observed D enhancement is well understood (e.g.\
Watson 1974; Gu\'elin \etal 1982; Millar, Bennett, \& Herbst 1989;
Roberts \& Millar 2000). Essentially, small zero-point energy
differences ensure that molecular ions become preferentially
deuterated via D/H exchange reactions with HD, and subsequent
ion-molecule reactions spread this D-enrichment to neutral species. At
10\,K, the most important fractionation process is that of H$_2$D$^+$:
\begin{equation}
  {\rm H_3^+ ~+~ HD ~~~\rightleftharpoons~~~ H_2D^+ ~+~ H_2} \label{Rfrac1}
\end{equation}
We can quantify the degree of fractionation attainable by
reaction~(\ref{Rfrac1}) by writing
\begin{equation}
{ R({\rm H_2D^+)} = {\cal S} R(\rm HD) }
\end{equation}
where $R$(HD) will be twice the cosmic D/H ratio of $\approx
1.6\times10^{-5}$ (Linsky \etal 1995), and ${\cal S}$ is the so-called
enhancement factor, given by:
\begin{equation}
  {\cal S} = { k_f \over k_r ~+~ \sum_j k_j x(m_j) ~+~
  \alpha_e({\rm H_2D^+}) x_e } \label{calS}
\end{equation}
(Stark, van der Tak, \& van Dishoeck 1999; Millar \etal 2000). $k_f$
and $k_r$ are the forward and reverse rate coefficients for reaction
(\ref{Rfrac1}); at 10\,K the respective values are $1.7\times10^{-9}$
and $3.6\times10^{-18}$ cm$^3$s$^{-1}$ (Millar \etal 1989; Sidhu,
Miller, \& Tennyson 1992), $x_e$ is the electron fraction, $k_j$ is the
rate coefficient for proton or deuteron transfer to species $m_j$
(principally CO, N$_2$ and O) whose abundance is $x(m_j) \equiv
n(m_j)/n(\rm H_2)$, and $\alpha_e(\rm H_2D^+)$ is the electron
dissociative recombination coefficient of $\rm H_2D^+$ (equal to $6
\times 10^{-8}$ ($T$/300\,K)$^{-0.65}$ cm$^{3}$s$^{-1}$; Larsson \etal
1996).

Proton transfer from H$_3^+$ to CO and N$_2$ is the primary source of
the HCO$^+$ and N$_2$H$^+$ ions, so the H$_2$D$^+$ enhancement also
determines the fractionation of DCO$^+$ and N$_2$D$^+$\@. If we assume
that one third of such reactions result in deuteron transfer, it
follows that
\begin{equation}
 R({\rm DCO^+}) = R({\rm N_2D^+}) = \case{1}{3} R({\rm H_2D^+}) =
 \case{1}{3}{\cal S}\,R({\rm HD}) \label{DCOP}
\end{equation}
This simple theory predicts that the fractionation of N$_2$D$^+$ and
DCO$^+$ should be equal. However, as discussed in $\S$1, the values
measured in L134N by Tin\'e \etal (2000) are 0.35 and 0.18
respectively. This discrepancy is puzzling, since if anything, the
value of $R$(DCO$^+$) should be larger than $R$(N$_2$D$^+$) as DCO$^+$
is also fractionated via the reaction of HCO$^+$ with atomic D (Adams
\& Smith 1985; cf.\ Dalgarno \& Lepp 1984; Opendak 1993). A possible
explanation is that deuteron transfer from H$_2$D$^+$ to N$_2$ occurs
preferentially to proton transfer, but that the branching ratios for
\mbox{$\rm H_2D^+ + CO$} are statistical. Nevertheless, the N$_2$D$^+$
fractionation is comparable to that determined by Snyder \etal (1977)
and, to order of magnitude, the observed fractionation is compatible
with the above theory if ${\cal S} \gtrsim 10^4$.


\section{Gas Phase Ammonia Formation}
\label{secgasamm}

The observed ammonia abundance in L134N is $\sim10^{-7}$ (Swade 1989;
Dickens \etal 2000), in agreement with the steady-state value in
chemical models where most of the nitrogen is in molecular form (e.g.\
Millar, Farquhar, \& Willacy 1997). In this case, ammonia formation is
initiated by He$^+$ attack on N$_2$ to form N$^+$\@. This ion then
undergoes successive reactions with H$_2$ until NH$_4^+$ is formed,
which then recombines to give \ammonia:
\begin{equation}
  {\rm N_2 ~\stackrel{He^+}{\longrightarrow}~ N^+
  ~\stackrel{4H_2}{\longrightarrow}~ NH_4^+
  ~\stackrel{\it e^-}{\longrightarrow}~ NH_3 } \label{seq}
\end{equation}
This formation mechanism remains somewhat controversial, because the
first hydrogenation step
\begin{equation}
{\rm N^+ + H_2 ~\longrightarrow~ NH^+ + H} \label{NplusH2}
\end{equation}
is endothermic by $\approx170$\,K (Marquette, Rebrion, \& Rowe 1988).
However, this is slightly less than the ground state rotational
energy of ortho-H$_2$, and Le Bourlot (1991) showed that as long as
the ortho-to-para ratio is greater than $\sim 10^{-4}$ then reaction
(\ref{NplusH2}) is the dominant loss route for N$^+$, so reaction
sequence (\ref{seq}) proceeds efficiently.  Even at 10\,K, chemical
cycling between ortho and para-H$_2$ via proton transfer reactions is
able to provide sufficient ortho-H$_2$ (Le Bourlot 1991, 2000).

At the densities appropriate for L134N, gas phase chemistry, rather
than freeze-out onto grains, is the dominant loss route for ammonia.
Cosmic ray ionization of H$_2$ followed by successive proton transfer
reactions results in NH$_3$ molecules being recycled into NH$_4^+$
ions, which can then recombine to either \ammonia\ or NH$_2$\@. In the
latter case, the NH$_2$ is rapidly destroyed by atomic oxygen. The
resulting expression for the steady-state ammonia abundance is:
\begin{equation}
 x(\ammonia) = {0.3\epsilon\zeta x({\rm He}) x({\rm N_2}) \over
 (1-\epsilon) [ x({\rm CO}) + x({\rm N_2}) ] \gamh }
 \label{ssamm}
\end{equation}
where $\epsilon$ is the fraction of $\rm NH_4^+$ dissociative
recombinations that lead to \ammonia, $\zeta$ is the cosmic ray
ionization rate, and \gamh\ is the total rate (s$^{-1}$) for proton
transfer to \ammonia. In theory, \gamh\ will depend on the abundances
of all molecular ions capable of transferring a proton to ammonia, but
because HCO$^+$, N$_2$H$^+$ and H$_3^+$ are the dominant ions we can
write
\begin{equation}
 \gamh \approx 10^{-9}\,[ 2.2n({\rm HCO^+}) +
 2.3n({\rm N_2H^+}) + 2.7n({\rm H_3^+}) ]~~~~~~ \label{Kp}
\end{equation}
Based on the fact that the abundances of HCO$^+$ and N$_2$H$^+$ are
observationally well-determined in L134N, and that the abundance of
H$_3^+$ can be obtained from simple chemical arguments (see
$\S\ref{secL134N}$), we derive $\gamh \approx 10^{-12}$~s$^{-1}$. The
value of $\epsilon$ was measured in the laboratory by Vikor \etal
(1999), and was found to be 0.69. With assumed values for $x({\rm He})
= 0.15$, and $\zeta = 1.3\times10^{-17}$~s$^{-1}$, the observed
ammonia abundance in L134N can be reproduced by equation~(\ref{ssamm})
if the N$_2$/CO ratio is $\approx 0.1$, consistent with the N$_2$
abundances determined in a number of dark clouds by Womack, Ziurys, \&
Wyckoff (1992b).


\section{Gas Phase Ammonia Fractionation }
\label{secgasfrac}

\begin{figure*}[t!]
\psfig{file=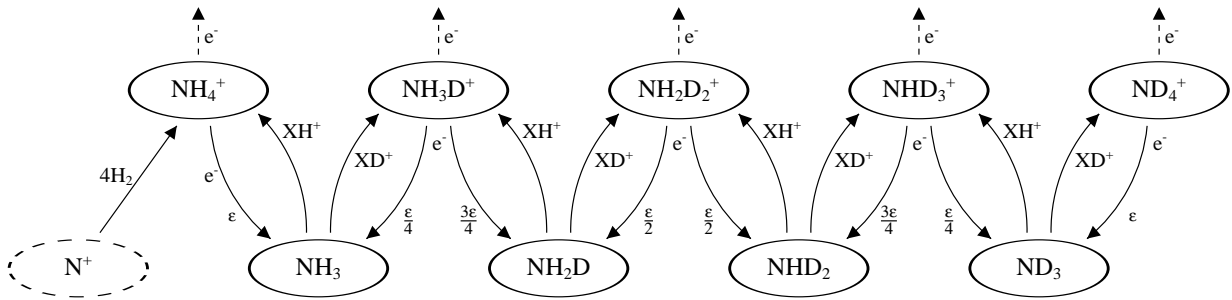,silent=,width=500pt}
\caption{Reaction network for ammonia deuteration in cold
 clouds. XH$^+$ and XD$^+$ represent all species capable of
 transferring a proton or deuteron to \ammonia; principally H$_3^+$,
 N$_2$H$^+$, HCO$^+$, and their deuterated isotopomers. Also shown are
 the assumed statistical branching ratios for each dissociative
 recombination channel. $\epsilon$ is the fraction of dissociative
 recombinations which result in an isotopomer of \ammonia\ rather than
 NH$_2$.}
\end{figure*}

We have extended our steady-state analysis to include multiply
deuterated ammonia. Figure~1 shows the primary chemical reactions
creating and destroying \ammonia\ and its associated isotopomers
through ion-molecule reactions involving generic protonated and
deuterated ions, XH$^+$ and XD$^+$\@.  After \ammonia\ is formed by
the reaction sequence (\ref{seq}), deuteron transfer reactions form
\pdamm\ which can then recombine to give \damm. Successive deuteron
transfer reactions can lead eventually to \ddamm\ and \tdamm. The
relative steady-state abundances therefore depend on the XD$^+$/XH$^+$
ratio and the branching ratios for dissociative recombination of the
deuterated ions.

We can quantify the XD$^+$/XH$^+$ ratio by introducing the parameter
$\bar{R}$, equal to the ratio of the rates for deuteron vs.\ proton
transfer, i.e.\ we define $\bar{R} \equiv \gamd/\gamh$, where \gamh\
is given by equation (\ref{Kp}) and \gamd\ is obtained from
\begin{equation}
 \gamd \approx 10^{-9}\,[ 2.2n({\rm DCO^+}) +
 2.3n({\rm N_2D^+}) + 0.9n({\rm H_2D^+}) ]~~~~~~ \label{Kd}
\end{equation}
As with \gamh, there will also be a small contribution to \gamd\ from
isotopomers of less abundant ions (e.g.\ H$_3$O$^+$, HOCO$^+$,
HCNH$^+$\@, etc.), but because HCO$^+$ is the most abundant ion, and
because the deuterium fractionation in these minor ions originates
with deuteron transfer from H$_2$D$^+$ and DCO$^+$, in practice
$\bar{R} \approx R({\rm DCO^+})$.

We assume that the \ammonia/NH$_2$ branching ratio, $\epsilon$, is the
same for deuterated ammonium ions as for \pamm, and we assume
statistical branching ratios regarding the position of the D in the
products (see Fig.~1). Gellene \& Porter (1984) measured the branching
ratios for electron dissociative recombination of NHD$_3^+$ and found
H atom ejection occurred 2.7 times more frequently than D ejection,
implying that the N--H bonds are eight times more likely to break than
the N--D bond. A similar effect has been observed in the recombination
of HDO$^+$ and HD$_2$O$^+$ (Jensen \etal 1999, 2000).  Therefore, our
results represent a lower limit to the amount of deuteration that can
occur via ion-molecule chemistry, since preferential retention of the
D in the molecule is more likely to occur.

For each of the species in Fig.~1, one can write an expression
equating the formation and destruction rates. These expressions can
then be solved to give the number density ratios at steady
state. After some algebra one obtains:
\begin{eqnarray}
 R(\tdamm) &=& { \epsilon\bar{R} \over
 (4-\epsilon) + (4-4\epsilon)\bar{R} } \label{r3}
\\
 R(\ddamm) &=& {2\epsilon\bar{R} \over
 (4-2\epsilon) + (4-3\epsilon)\bar{R} - 3\epsilon R({\rm ND_3})}
 \label{r2}
\\
 R(\damm) &=& {3\epsilon\bar{R} \over
 (4-3\epsilon) + (4-2\epsilon)\bar{R} - 2\epsilon R({\rm NHD_2})}
 \label{r1}
\end{eqnarray}
Note that, although the {\it absolute} abundances will depend on the
formation rate of \ammonia, the {\it relative} abundances of the
deuterated forms depend only on $\bar{R}$ and $\epsilon$, and so are
not affected by any uncertainty regarding the kinetics of ammonia
synthesis.

\begin{figure*}
\psfig{file=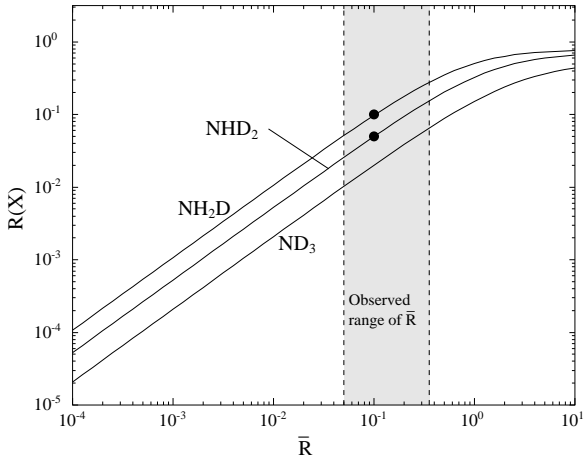,silent=,width=235pt}
\caption{Steady-state fractionation ratios for deuterated ammonia
 isotopomers, as a function of the mean deuterium fractionation in
 molecular ions, $\bar{R}$\@. The curves correspond to solutions of
 equations (\protect\ref{r3})--(\protect\ref{r1}) for $\epsilon =
 0.69$, the filled circles show the observed values of $R(\damm)$ and
 $R(\ddamm)$, and the shaded region illustrates the range of values of
 $\bar{R}$ implied by different observations.}
\end{figure*}

Solutions of equations (\ref{r3})--(\ref{r1}) are plotted in Figure~2.
For $\bar{R} \lesssim 1$ the predicted values are proportional to
$\bar{R}$\@. At larger values of $\bar{R}$ the curves asymptotically
approach the value determined by the branching ratios of the
deuterated ammonium ions. The shaded region in Fig.~2 covers the
observed range in $\bar{R}$; from a lower bound of 0.05 inferred from
DCO$^+$ observations (see $\S 1$), up to a maximum of 0.35 implied by
the N$_2$D$^+$ observations of Tin\'e \etal (2000). The filled circles
in Fig.~2 show the observed \damm\ and \ddamm\ ratios; the fact that
both the observed values imply the {\it same} underlying value of
$\bar{R}$ is convincing evidence that gas-phase chemistry is the
source of the \damm\ and \ddamm\ in L134N\@.

Note that there also exists another gas phase route to
singly-deuterated ammonia: the reaction of N$^+$ with HD, analogous to
reaction~(\ref{NplusH2}). In this case, the channel leading to $\rm
ND^+ + H$ is favored over $\rm NH^+ + D$, since the endothermicity of
the former channel is only 16\,K (Marquette \etal 1988). Hence, if the
H$_2$ ortho-to-para ratio is low so that reaction~(\ref{NplusH2})
occurs slowly, this can lead to large enhancements of \damm\ via the
deuterated version of reaction sequence~(\ref{seq}) (Tin\'e \etal
2000). However, this mechanism cannot explain the presence of
doubly-deuterated ammonia. As we have shown that deuteron transfer
reactions are more than capable of producing the observed \damm\ and
\ddamm\ abundances, we believe that the scheme illustrated in Fig.~1
is sufficient to explain the ammonia deuteration in L134N\@.


\section{The Chemical State of L134N}
\label{secL134N}

We have shown that the observed abundances of ammonia isotopomers in
L134N are consistent with the steady-state solutions of a relatively
simple ion-molecule reaction scheme (see Fig.~1), if the value of
$\bar{R}\approx 0.1$. In order to assess the validity of our analysis,
we would like to be able to reconcile our scheme with the observed
abundances of simple molecular ions and their deuterated counterparts.

The observed range of values for $\bar{R}$ implies a value of $\cal S$
of 5000--35000 (eqn.~[\ref{DCOP}]). In order to calculate the expected
value of $\cal S$ in L134N from equation~(\ref{calS}), we need to know
the electron fraction, $x_e$, and $\sum_j k_j x(m_j)$, the total
removal rate of H$_2$D$^+$ through reactions with heavy species. We
can calculate these values approximately from the the steady-state
H$_3^+$ concentration, which is given by
\begin{equation}
  n({\rm H_3^+})= { {\zeta} \over { \sum_j k_j x(m_j) ~+~
   \alpha_e({\rm H_3^+}) x_e ~+~ k_f R({\rm HD}) } }
  \label{h3plus}
\end{equation}
and the charge conservation equation:
\begin{equation}
  x_e \approx x({\rm HCO^+}) ~+~ x({\rm N_2H^+}) ~+~ x({\rm H_3^+})
  \label{econs}
\end{equation}
Note that the rate coefficients in equation~(\ref{h3plus}) are the
same as in equation~(\ref{calS}), since we assume that H$_2$D$^+$
reacts at the same rate as H$_3^+$, except for electron recombination,
where we use the value for $\alpha_e({\rm H_3^+})$ of
$1.15\times10^{-7}(T/300\,{\rm K})^{-0.65}$ measured by Sundstr\"om
\etal (1994).

HCO$^+$ and N$_2$H$^+$ have been observed in L134N, and their
respective abundances are $1.2\times10^{-8}$ and $7\times10^{-10}$
(Swade 1989; Womack \etal 1992a; Dickens \etal 2000). Adopting
physical parameters of $T=10$\,K and $n({\rm H_2}) =
2\times10^4$~cm$^{-3}$ (Dickens \etal 2000), we thus have three
equations ([\ref{calS}], [\ref{h3plus}], and [\ref{econs}]) in three
unknowns ($x_e$, $x({\rm H_3^+})$, and $\sum_j k_j x(m_j)$).  Solving
these equations for different values of $\cal S$ allows us to
constrain the chemical state of L134N\@. Also, because the rate
coefficients for reactions of H$_3^+$ and H$_2$D$^+$ with heavy
molecules are $\approx 2\times10^{-9}$~cm$^3$~s$^{-1}$, we can
calculate the total abundance of heavy molecules from our derived
value of $\sum_j k_j x(m_j)$.

For ${\cal S} = 5000$, we calculate $x_e = 1.4\times10^{-8}$, $x({\rm
H_3^+}) = 1.6\times10^{-9}$, and $\sum_j x(m_j) = 1.6\times10^{-4}$,
consistent with little depletion of CO, N$_2$, and O from the gas
phase. On the other hand, for ${\cal S} = 35000$, we derive $x_e =
1.9\times10^{-8}$, $x({\rm H_3^+}) = 6\times10^{-9}$, and $\sum_j
x(m_j) = 1.9\times10^{-5}$, implying significant depletion. For the
intermediate value of ${\cal S} \approx 10^4$ inferred from the
deuterated ammonia fractionation, we find that partial depletion
($\sim 50$\%) of heavy molecules is required to account for the
observed deuterium enhancements. A similar conclusion has been reached
by Roberts \& Millar (2000). Finally, as our calculated ionization
levels lead to good agreement with the observed \ammonia\ abundance
($\S\ref{secgasamm}$), we conclude that the observed abundances and
fractionations in L134N are well-matched by steady-state ion-molecule
chemistry.


\section{Gas Phase vs. Grain Surface Ammonia Fractionation}

Despite reservations concerning the physical conditions in L134N, some
as yet unidentified mechanism may be returning mantle-formed molecules
to the gas there. The mantle abundances of ammonia isotopomers
computed numerically by Brown \& Millar (1989) (see their table 2)
effectively rule out a grain surface origin when scaled to the ammonia
abundance in L134N\@. However, their surface reaction scheme permits
larger D/H ratios than those presented, since the fractionation is
proportional to the gas phase atomic D/H ratio, $R$(D), which may be
higher than the value assumed in their calculations. Following the
scheme of Brown \& Millar, we can derive values for the surface
fractionation ratios:
\begin{eqnarray}
 R_s(\damm) &~=~& \case{3}{\sqrt 2} R({\rm D}) \label{surf1}
\\
 R_s(\ddamm) &~=~& \case{1}{\sqrt 2} R({\rm D})
\\
 R_s(\tdamm) &~=~& \case{1}{3\sqrt 2} R({\rm D}) \label{surf3}
\end{eqnarray}
Thus, if $R({\rm D}) \approx 0.05$, then it is also possible to
explain the observed abundances by surface formation of ammonia.

It may be possible, however, to discriminate between alternative
formation mechanisms by examining the relative scaling of the
fractionation ratios. For example, grain surface formation implies
that
\begin{equation}
 R_s(\damm):R_s(\ddamm):R_s(\tdamm) =
 1 : 0.33 : 0.11 \label{surfscale}
\end{equation}
whereas, to first order, gas phase chemistry implies
(using equations [\ref{r3}]--[\ref{r1}])
\begin{equation}
 R(\damm):R(\ddamm):R(\tdamm) =
 1 : 0.49 : 0.19 \label{gasscale}
\end{equation}
The fact that the calculated values of $R$ for the three isotopomers
always have the same ratio irrespective of their absolute values is
apparent from Fig.~2, where the vertical separation of the curves for
the three isotopomers is constant.

For doubly-deuterated ammonia, the predicted ratios are similar, with
the observed value of $R(\damm) = 0.1$ implying respective values of
$R(\ddamm) = 0.05$ and 0.03 for gas phase and surface chemistry.
Although the observed value of 0.05 agrees with our gas phase scheme,
the observational uncertainties are too large to rule out surface
formation. However, for triply-deuterated ammonia, the gas phase
scheme predicts a fractionation almost twice as large; when scaled to
the \ammonia\ abundance this implies a \tdamm/\ammonia\ ratio 2.5
times greater than the value predicted by surface chemistry (where we
have used the relation $\tdamm/\ammonia = R(\tdamm)\times
R(\ddamm)\times R(\damm)$; cf.\ our definition of $R$ in $\S
1$)\@. With the observed values $x(\ammonia) = 10^{-7}$ and $R(\damm)
= 0.1$, we predict $x(\tdamm) \approx 10^{-11}$. Hence, if \tdamm\ can
be detected (or an upper limit determined) in L134N, it may be
possible to determine whether the ammonia is formed in the gas or on
the grains.

It is worth stressing that the kind of scaling relations for
multiply-deuterated fractionation ratios expressed by
equations~(\ref{surfscale}) and (\ref{gasscale}) are applicable in
general to all molecules. Thus, whereas the fractionation of
singly-deuterated molecules reflects both the formation mechanism of
the molecule and the underlying D/H ratio in the precursors, the
relative fractionation ratios of multiply-deuterated molecules reflect
only the formation mechanism. This fact was first appreciated by
Turner (1990), who used the D$_2$CO:HDCO:H$_2$CO ratios to show that
formaldehyde in the Orion Compact Ridge should have a grain surface
origin.


\section{Discussion}
\label{seclast}

We have shown that large abundances of \damm\ and \ddamm\ can be
produced by gas phase chemistry in the interiors of cold dense
clouds. Ammonia is deuterated via deuteron transfer from species
such as H$_2$D$^+$, DCO$^+$, and N$_2$D$^+$, followed by dissociative
recombination. This mechanism is able to match the observed
fractionation ratios of both species if the underlying XD$^+$/XH$^+$
ratio, $\bar{R}$, equals 0.1.

Grain surface formation of ammonia produces distinct fractionation
ratios, however the uncertainties in the observed abundances mean that
we cannot definitively conclude that deuterated \ammonia\ is being
formed in the gas. Because the scaling of the fractionation ratios
expected from these two processes is not the same, this raises the
possibility that the \damm:\ddamm:\tdamm\ ratios may ultimately be
used to determine the origin of these molecules.  In particular, this
could be resolved with the detection of \tdamm, which we predict to
have an abundance of $\sim 10^{-11}$ in L134N\@. Gas phase formation
appears more feasible than surface chemistry since it is able to
account for the observed \ammonia\ abundance without recourse to
uncertain surface processes and desorption mechanisms. A further
problem for the grain surface hypothesis is the fact that the large
value of $R$(\damm) in L134N requires a gas phase atomic D/H ratio of
0.05, but theoretical models predict an equilibrium value of only a
few times $10^{-3}$ at 10\,K (Millar \etal 1989; Roberts \& Millar
2000).

We also find that the large observed molecular D/H ratios can only be
reproduced if heavy elements are partially depleted onto grain
surfaces. Therefore, it appears that the deuterium emission peak in
L134N traces a small region where significant amounts of CO, N$_2$ and
O are frozen onto grains. A similar region of enhanced D fractionation
is known to exist in TMC-1 (Gu\'elin \etal 1982), a dark cloud that
appears to be physically similar to L134N\@.  The fact that there
appear to be several infrared sources located behind L134N (Snell
1981) may allow the molecular depletion into ice mantles to be
measured and crudely mapped. Alternatively, these spatial gradients
may be due to the energy available when ions and neutrals have
slightly different velocities. In this case, $k_r$ becomes the
dominant term in the denominator of the expression for ${\cal S}$
(eqn.~[\ref{calS}]), which is consequently reduced (Charnley 1998).
Another explanation may be the existence of chemical bistability in
interstellar clouds; Gerin \etal (1997) showed that molecular D/H
ratios are typically reduced by a factor of ten in the high ionization
phase steady-state solution, as opposed to the low ionization phase.

Nevertheless, it seems more likely that depletion is the cause of the
high deuteration in L134N, since the latter mechanisms act to reduce
$\cal S$ whereas depletion causes $\cal S$ to increase. It is
interesting to note that the observed range of $\bar{R}$ in L134N is
always $\gtrsim 0.05$; this is what one would expect for a 10\,K cloud
with no depletion, and is in fact the value derived from DCO$^+$
observations of a large number of cold clouds (Gu\'elin \etal 1982;
Butner \etal 1995). The fact that the fractionation in L134N is
determined to be above this canonical cold cloud value is evidence for
selective deuterium enhancement in this particular region, as opposed
to a reduction of the fractionation in the surrounding gas. On the
other hand, Gerin \etal (1997) observed a value of $R({\rm DCO^+}) =
0.003$ in the high latitude cloud MCLD~123.5$+$24.9, ten times lower
than the `normal' value. In this instance, it would appear that some
mechanism is indeed operating to suppress the D/H enhancement.

If the depletion is particularly large, the value of $\bar{R}$ can
become as large as unity (Millar \etal 2000). In this case, chemical
models predict large abundances of HDCO and D$_2$CO (Roberts \& Millar
2000), so it may be fruitful to search for D$_2$CO in dark
clouds. Note, however, that deuterated formaldehyde may not in fact be
as abundant as predicted by these models where it is assumed that
H$_2$CO is deuterated via the same mechanism as \ammonia, namely
deuteron transfer from H$_2$D$^+$, DCO$^+$ etc.\ followed by
dissociative recombination. In fact, the lowest energy isomer of
protonated formaldehyde has the form H$_2$COH$^+$, and this ion has
been detected in the interstellar medium (Ohishi \etal 1996). We may
therefore expect that deuteron transfer to H$_2$CO will most likely
result in H$_2$COD$^+$ and will not lead to HDCO after dissociative
recombination (Sen, Anicich, \& Federman 1992).

Finally, we note that the DCO$^+$ fractionation of 0.05 observed in a
number of sources by Butner \etal (1995) is only a factor of two less
than the value of $\bar{R}$ required to explain the ammonia
deuteration in L134N\@. Therefore, multi-deuterated ammonia may be
widespread in dark clouds.


\acknowledgements{Theoretical astrochemistry at NASA Ames is supported
by NASA's Origins of Solar Systems and Exobiology Programs through
NASA Ames Interchange NCC2-1162. S. D. R. is supported by a National
Research Council postdoctoral research associateship.}


\end{document}